\documentclass[10pt]{article}
\usepackage{graphicx}
\usepackage{amsfonts}

\begin{document}

\newcommand {\lum} {\, \mbox{${\rm cm}^{-2} {\rm s}^{-1}$}}
\newcommand {\barn} {\, \mbox{${\rm barn}$}}
\newcommand {\m} {\, \mbox{${\rm m}$}}
\newcommand {\bstar} {\, \mbox{$\beta^{*}$}}
\newcommand{\pt}{\ensuremath{p_{\mathrm{t}}} }
\newcommand{\dd}{\ensuremath{\mathrm{d}}}
\newcommand{\Jpsi}    {\mbox{J\kern-0.05em /\kern-0.05em$\psi$}}
\newcommand{\psip}    {\mbox{$\psi^\prime$}}
\newcommand{\Ups}     {\mbox{$\Upsilon$}}
\newcommand{\Upsp}    {\mbox{$\Upsilon^\prime$}}
\newcommand{\Upspp}   {\mbox{$\Upsilon^{\prime\prime}$}}
\newcommand{\qqbar}   {\mbox{$q\bar{q}$}}
\newcommand{\pT}      {\ensuremath{p_{\mathrm{t}}}}
\newcommand{\mev}     {\mbox{${\rm MeV}$}}
\newcommand{\gev}     {\mbox{${\rm GeV}$}}
\newcommand{\pTpa}    {\langle \pT ^ {\rm parton} \rangle}
\newcommand{\pTp}     {\pT ^{\rm parton}}
\newcommand{\pTt}     {\pT^{\rm trig}}
\newcommand{\pTa}     {\pT^{\rm assoc}}
\newcommand{\zT}      {\ensuremath{z_{\mathrm{T}}}}

\title{On the Mean Parton Transverse Momentum versus Associated Hadron $\pT$ 
in Di-Hadron Correlations at RHIC and LHC}
\author{A. Morsch}

\maketitle

\begin{abstract}
From Pythia simulations of pp collisions we determine the mean transverse 
momentum $\pTpa$ of partons 
producing the leading and associated particles studied in di-hadron correlations. 
The dependence of $\pTpa$ on the associated particle transverse momentum and 
expected differences at RHIC and LHC centre of mass energies are discussed.
\end{abstract}

\section{Introduction} 

Azimuthal correlations of high-$\pT$ hadrons are used at RHIC to study the fragmentation properties
of partons produced in $2 \rightarrow 2$  hard scatterings. The correlations are studied as 
a function of the transverse momentum of the trigger hadron $\pTt$ and the associated particle
$\pTa$ \cite{Adler:2002tq, Adler:2003qi}. 
The angular correlation functions show two characteristic approximately Gaussian peaks for hadrons 
produced from the fragmentation of the same parton that produced the trigger hadron (near-side) peak and 
from the fragmentation of the partons produced back-to-back approximately balancing the transverse momentum of
the trigger parton (away-side peak). 
A strong suppression of the back-side peak in central Au+Au collisions relative to pp collisions
has been observed and interpreted as evidence for in-medium partonic energy loss.

It has been demonstrated that due to the bias induced by the steeply falling production spectrum 
${\rm d}\sigma/{\rm d} \pTp$, high-$\pT$ hadrons carry about $60-70\%$ of the original
parton transverse momentum (see for example \cite{Wang:2003aw}).
Due to the steepness of the partonic fragmentation  function $D(z)$ with $z = \pTa/\pTp$,
it can be expected that demanding the presence of an additional associated 
particle further increases the parton mean transverse momentum $\pTpa$ (\cite{Adler:2006sc}): 
for constant $\pTa$, $z$ decreases with increasing $\pTp$.
This effect should be stronger for the near-side associated hadrons, 
owing to the trivial fact that $\pTp$ can not be smaller than the sum $\pTt + \pTa$.
The on average higher $\pTp$ and the corresponding smaller production cross-section
have to be taken in account when comparing near-side to away-side 
associated hadron $\pT$ spectra and widths of angular correlations.

In the next sections we show the results of Pythia simulations establishing the relation between  
$\pTp$ and the associated parton transverse momentum in the range $\pTa > 0 - 8 \, \gev $ 
for $\pTt > 8 \, \gev$ at RHIC and LHC energies. 
The choice of cuts may allow to use these results for further interpretation of the 
di-hadron correlation data recently published by the STAR collaboration \cite{Adams:2006yt}.

\begin{figure}[htb]
  \begin{center}
    \includegraphics[width=.48\textwidth]{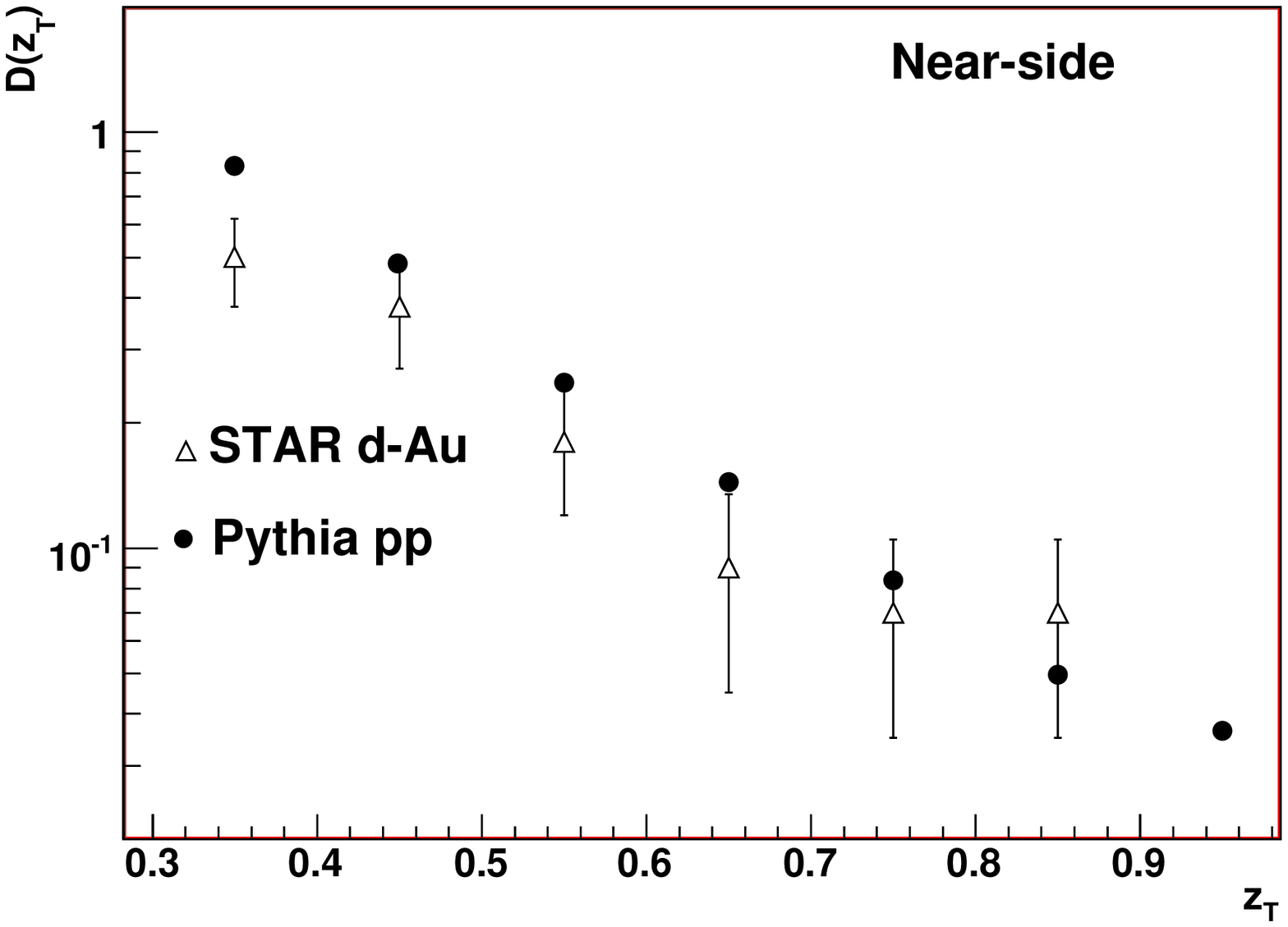}
    \includegraphics[width=.48\textwidth]{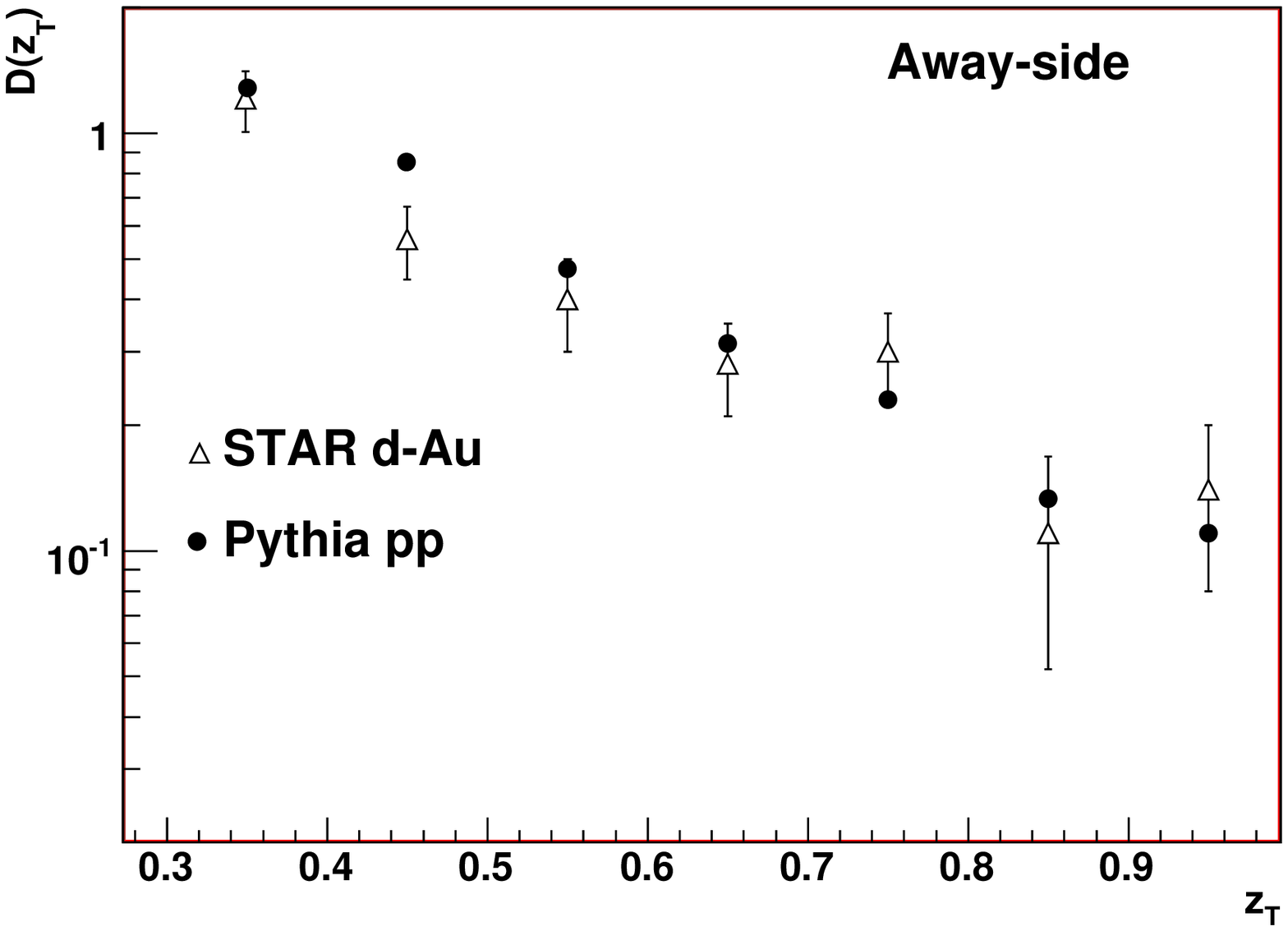}
    \caption{
      Trigger normalised charged hadron fragmentation function $D(\zT)$ with 
      $\pTt > 8 \, \gev$ obtained from a Pythia simulation of pp collisions at 
      $\sqrt{s} = 200 \, \gev$ as described in the text. 
    The simulation results are compared with the STAR d-Au data 
    \cite{Adams:2006yt} (data values read from the figure).}
      \label{fig:frag}
  \end{center}
\end{figure}

\begin{figure}[htb]
  \begin{center}
    \includegraphics[width=.95\textwidth]{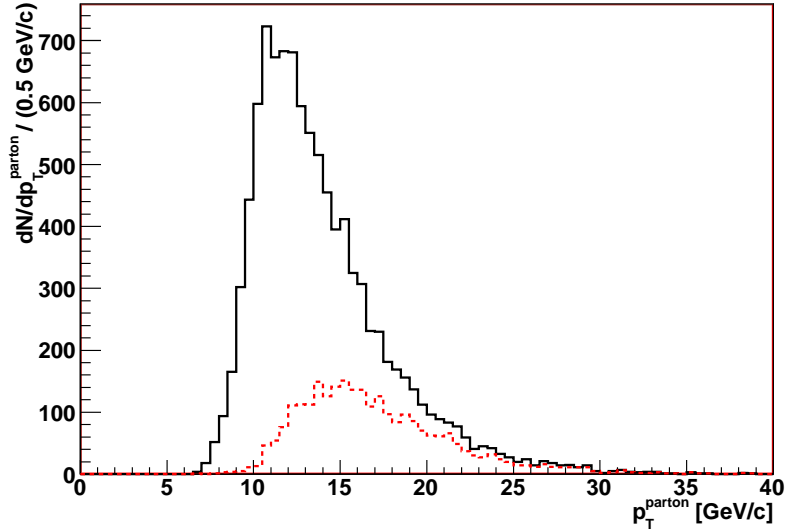}
    \caption{
      The parton transverse momentum spectrum selected by 
      $\pTt > 8 \, \gev , \, \pTa > 0 \, \gev$ (solid line) 
      is compared to the one selected by
      $\pTt > 8 \, \gev , \, \pTa > 2 \, \gev$. (dashed line, online:red).
      The distribution corresponds to a statistics of $10^4$ triggers.
    }
    \label{fig:res1}
  \end{center}
\end{figure}

\begin{figure}[htb]
  \begin{center}
    \includegraphics[width=.95\textwidth]{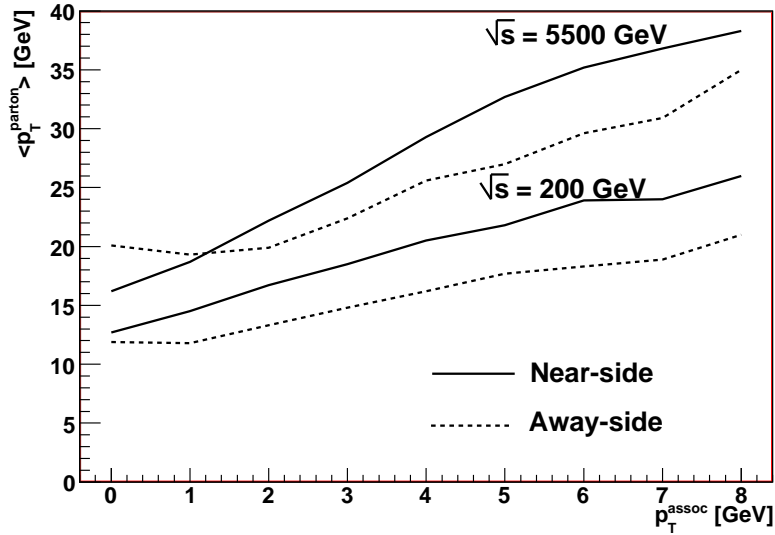}
    \caption{
      Average parton transverse momentum ($\pTp$) as a function of the associated hadron transverse momentum
      for $\pTt > 8 \, \gev$ in pp collisions at $\sqrt{s} = 200 \, \gev$ and $\sqrt{s} = 5500 \, \gev$.
    }
    \label{fig:res2}
  \end{center}
\end{figure}

\begin{figure}[htb]
  \begin{center}
    \includegraphics[width=.95\textwidth]{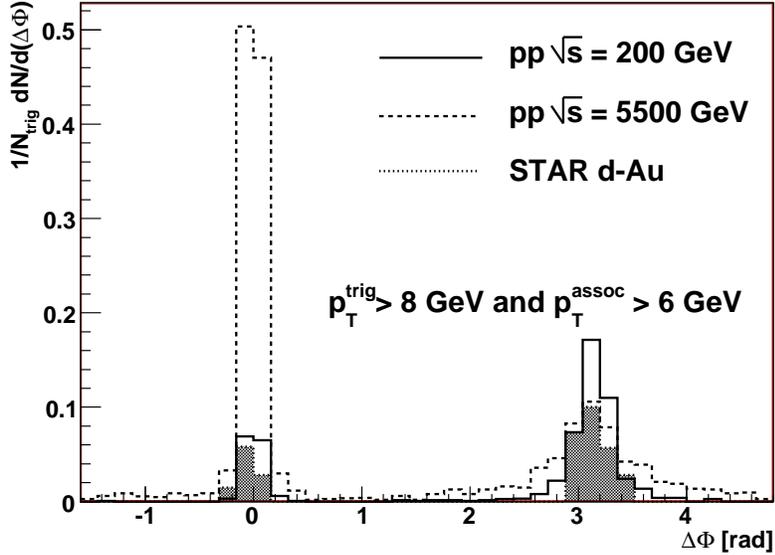}
    \caption{
      Azimuthal correlation function for $\pTt > 8 \, \gev$, $\pTa > 6 \, \gev$ and $|\eta < 1|$  
      for pp collisions at $\sqrt{s} = 200 \, \gev$ (solid line) and $\sqrt{s}  = 5500 \, \gev $ (dashed line).
    }
    \label{fig:dphi}
  \end{center}
\end{figure}

\begin{figure}[htb]
  \begin{center}
    \includegraphics[width=.48\textwidth]{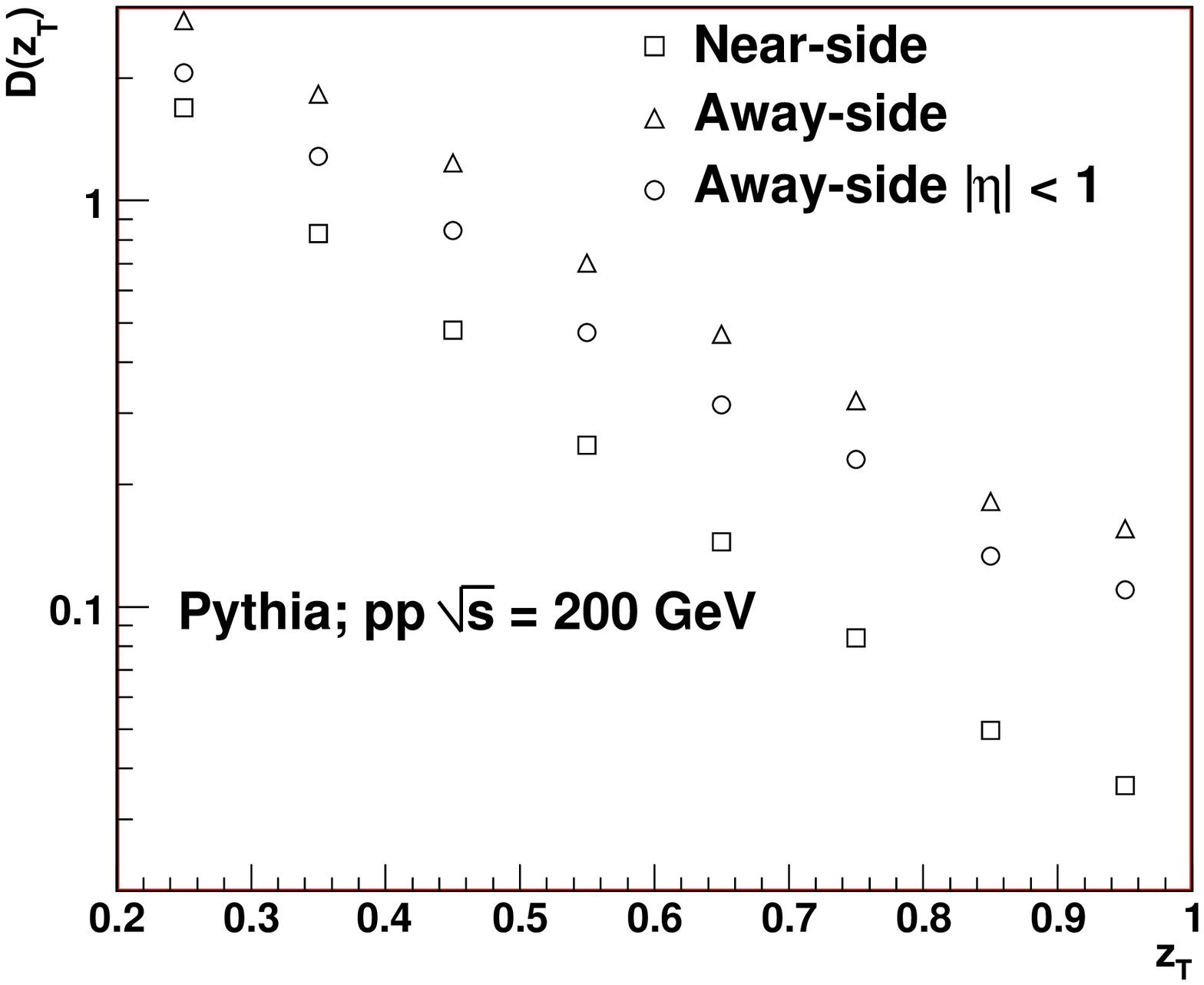}
    \includegraphics[width=.48\textwidth]{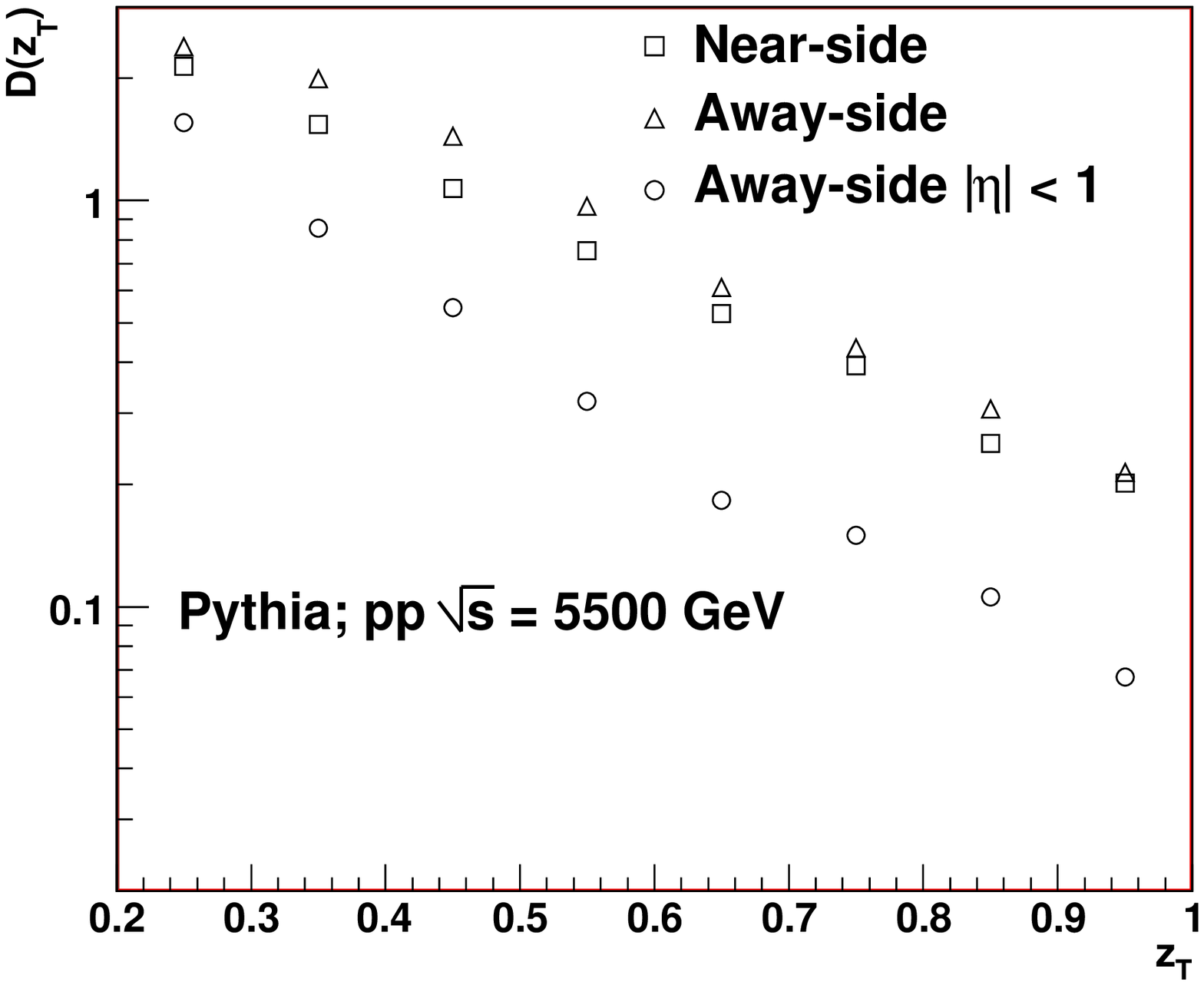}
    \caption{
      Trigger normalised charged hadron fragmentation function $D(\zT)$ with 
      $\pTt > 8 \, \gev$ obtained from a Pythia simulation of pp collisions at 
      $\sqrt{s} = 200 \, \gev$ (left) and $\sqrt{s} = 5500 \, \gev$ (right).
}
      \label{fig:frag_lhc}
  \end{center}
\end{figure}

\section{Method}
Pythia 6.2 \cite{Sjostrand:1985xi,Sjostrand:1987su} is used to simulate pp collisions at $\sqrt{s} = 200 \, \gev$
and $5500 \, \gev$ (QCD processes, MSEL=1). 
The transverse momenta of all final state charged hadrons and the partons resulting from the 
hard $2 \rightarrow 2$ scattering process are recorded. 
We select the hadron with the highest $\pT$, $\pT^{\rm max}$ within the pseudo rapidity window
$|\eta| < 1$, the trigger hadron.
If $\pT^{\rm max} > \pTt$ the event is selected.
Subsequently we identify the highest $\pTa$ hadron within a 
distance $\Delta \phi < 0.63$ to the trigger hadron, the associated near-side hadron and
within $|\Delta \phi - \pi| < 0.63$, the associated away-side hadron.

The associated parton transverse momentum 
is determined by identifying the parton emerging from the hard scattering closest 
in $\eta - \phi$ to the trigger particle. Alternatively we used the 
Pythia {\tt PYCELL} cone jet reconstruction algorithm (with $R=1$) 
to find the jet with the axis closest to the trigger hadron. Both methods give similar results.
The average parton momentum $\pTp$ is determined for different 
($\pTt ,\,  \pTa $)-cuts.

As a benchmark for our simulations we show in Fig. \ref{fig:frag}  the 
$D(\zT)$ ($\zT = \pTa/\pTt$) distributions for near-side and away-side correlated particles. 
The distributions agree very well with the STAR d-Au data \cite{Adams:2006yt}.

\section{Discussion of the results}
Fig. \ref{fig:res1} compares the  parton spectra selected by 
$\pTt > 8 \, \gev$ and  $\pTa > 0 \, \gev$ 
to the one selected by
$\pTt > 8 \, \gev$ and $\pTa > 2 \, \gev$. 
Already at this moderate associated hadron $\pT$ an important shift to higher $\pTp$ and 
a sizable reduction of rate is observed.

Fig. \ref{fig:res2} shows $\pTpa$ as a function of $\pTa$.
$\pTpa$ rises approximately linearly in the range $0 < \pTa < \pTt$. 
At RHIC the difference in $\pTpa$ for near-side and away-side associated partons reaches 
up $30\%$ at high $\pTa$. At LHC energies the difference is about a factor of $1.5$ smaller and 
$\pTpa$ is a factor of 1.5 higher. The difference is due to the different shape of the 
partonic production cross-section as a function of $\pTp$; 
${\rm d}\sigma/{\rm d} \pTp$ follows a power law $1/(p_{\rm T}^{\rm parton}) ^n$, where for parton energies considered here,
$n \approx 8$ at RHIC and $\approx 4$ at LHC. 

In a recent paper \cite{Adams:2006yt} the STAR collaboration reports on the direct observation of dijets 
in central Au+Au collisions at $\sqrt{s_{NN}} = 200 \, \gev$. 
The transverse momentum distributions of near- and away-side associated hadrons is used 
to study the effect of the medium on dijet fragmentation. 
The authors emphasize that  $D(\zT)$ is measurable without direct knowledge of the parton energy. 
Although this is true, one has to keep in mind when comparing near-side and away-side associated 
hadrons, that $\zT$ covers different parton energy regions. 

One directly observable effect is the kinematic suppression of the nearside peak seen in the STAR d-Au data.
For increasing $\pTa$, the amplitude of the near-side peak decreases relative to the away-side peak. 
The azimuthal correlation functions for $\pTt > 8 \, \gev$ and $\pTa > 6 \, \gev$ as obtained from the 
Pythia simulations are shown in Fig. \ref{fig:dphi}. 
Comparing the amplitude of the peaks 
at $\sqrt{s} = \, 200 \, \gev$ to $\sqrt{s} = 5500 \, \gev$ one observes
that the near-side peak rises by a factor of $\approx 7$, whereas the away-side peak decreases by a factor of $\approx 1.3$. 
 
The rise of the near-side peak is mainly due to the difference in the partonic production spectra for the two different energies.
Since the difference between near-side and away-side $\pTpa$ is smaller at LHC the rise of the 
away-side peak due to the same effect is smaller. 
In addition, there are at least two effects that are expected to lead to the suppression of the away-side peak at LHC energies relative to RHIC.
\begin{itemize}
\item Broadening of the pseudorapidity correlations due to the smaller Bjorken $x_B \sim 1/\sqrt{s}$ 
leads to a reduction by a factor of $2.4$ for $|\eta| < 1|$.
\item Broadening of the azimuthal correlations due to the increased contribution from higher order QCD processes.
The width of the away-side peak is by a factor of two larger at LHC energies.
\end{itemize}

The increased nearside correlation at higher centre of mass energies can already be seen in data 
on charged particle correlations published by the CDF collaboration \cite{Affolder:2001xt} for $\sqrt{s} = \, 1800 \,  \gev$
\footnote{In simulation, the forward-backward peak inversion 
between RHIC and LHC energies has first been remarked in a HIJING study
\cite{ploskon}}.  

The systematics of the ratios between near-side to away-side peaks for different $\pTa$ is obtained 
by comparing the corresponding fragmentation functions $D(\zT)$ (Fig. \ref{fig:frag_lhc}).

\section{Conclusions}
From Pythia simulations of pp collisions at RHIC and LHC energies 
we have determined the mean transverse momentum $\pTpa$
of partons fragmenting into a leading trigger hadron with transverse 
momentum $\pTt$ and/or an associated hadron with transverse momentum $\pTa$ for 
$\pTt > 8 \, \gev$ and $0 < \pTa < \pTt$.
For both, near-side and away-side correlations,
$\pTp$ depends not only on the $\pTt$ but also strongly on the associated hadron $\pTa$.
Due to the additional kinematic constraint $\pTp > \pTt + \pTa$, 
$\pTpa$ is higher for near-side correlated hadrons than for away-side correlations.
The difference amounts to up to 30\% at RHIC and 20\% at LHC.

The higher $\pTpa$ and the corresponding smaller production cross-section
have to be taken in account when comparing near-side to away-side 
associated hadron transverse momentum spectra and widths of angular correlations.
In particular, at LHC energies, where the partonic production cross-section ${\rm d}\sigma/{\rm d} \pTp$
is less steep than at RHIC energies, the near-side azimuthal correlation 
peak will be enhanced with respect to the one observed at RHIC.

\section*{Acknowledgements}
The author greatly appreciates the fruitful discussions with G. Paic.


\begin{thebibliography}{1}
\bibitem{Adler:2002tq} C.~Adler {\it et al.} (STAR Collaboration),  Phys.\ Rev.\ Lett.\ {\bf 90} (2003) 082302.
\bibitem{Adler:2003qi} S.~S.~Adler {\it et al.} (PHENIX Collaboration), Phys.\ Rev.\ Lett.\ {\bf 91}  (2003) 072301.
\bibitem{Wang:2003aw}  X.~N.~Wang, Phys.\ Lett.\ B {\bf 579}, (2004) 299.
\bibitem{Adler:2006sc} S.~S.~Adler {\it et al.} (PHENIX Collaboration),  hep-ex/0605039.
\bibitem{Adams:2006yt} J. Adams {\it et al.} (STAR collaboration), nucl-ex/0604018.
\bibitem{Sjostrand:1985xi} T.~Sjostrand, Phys.\ Lett.\ B{\bf 157} (1985) 321.
\bibitem{Bengtsson:1986gz} M.~Bengtsson, T.~Sjostrand, and M.~van Zijl, Z.\ Phys.\ {\rm C32} (1986) 67.
\bibitem{Sjostrand:1987su} T.~Sjostrand and  M.~van Zijl, Phys.\ Rev.\ D{\bf 36}, (1987) 2019.
\bibitem{ploskon} H.~Appelsh\"auser and M.~Ploskon, ALICE Internal Note ALICE-INT-2005-49.
\bibitem{Affolder:2001xt} T.~Affolder {\it et al.} (CDF Collaboration) Phys.\ Rev.\ D{\bf 65} (2002) 092002. 

\end{thebibliography}
\end{document}